\documentclass[preprintnumbers,prd,nofootinbib,twocolumn,superscriptaddress, reprint]{revtex4-1}
\usepackage{amsmath,amssymb,bm,bbm,amsfonts}
\usepackage{physics}
\usepackage{graphicx,graphics}
\usepackage[dvipsnames]{xcolor}
\usepackage[colorlinks=true,linkcolor=blue,citecolor=blue,urlcolor=blue]{hyperref}
\usepackage{dsfont}
\usepackage{comment}
\usepackage{hhline,multirow}
\usepackage{dcolumn}
\usepackage{url}
\usepackage[normalem]{ulem}
\usepackage{mathrsfs}
\usepackage{latexsym}
\usepackage{amscd}
\usepackage{slashed}
\usepackage{mathtools}
\usepackage{cleveref}
\usepackage{float}

\usepackage[compat=1.1.0]{tikz-feynman}
\usepackage{tikz}
\usetikzlibrary{arrows.meta}
\usetikzlibrary{angles,quotes}
\usetikzlibrary {shapes.symbols}

\def\be{\begin{equation}}
\def\ee{\end{equation}}
\def\bea{\begin{eqnarray}}
\def\eea{\end{eqnarray}}

\begin{document}

\preprint{ADP-25-19/T1281}


\title{Relaxing constraints on a broad dark photon}

\author{J. R. Felix}
\author{A. W. Thomas}
\author{X.G. Wang}
\affiliation{CSSM and ARC Centre of Excellence for Dark Matter Particle Physics, Department of Physics, University of Adelaide, Adelaide, SA 5005, Australia}

\begin{abstract} 
We revisit the exclusion constraints on the parameters of a narrow dark photon set by direct experimental searches. We investigate how a dark photon with a larger decay width impacts these limits, in particular, in the case where the dark photon also decays into light dark matter.
As an example, taking the upper limits on the mixing parameter, $\epsilon$,  reported by the CMS collaboration, we find that they could be significantly relaxed. Indeed, even a very modest coupling of the dark photon to dark matter can lead to an increase in the bound on the mixing parameter by an order of magnitude.
\end{abstract}

\maketitle


\section{Introduction}
\label{sec:Intro}

Despite the great success of the Standard Model (SM), many potential phenomena suggest the existence of new physics beyond the SM, such as anomalies in the muon $g-2$~\cite{Muong-2:2021ojo, Muong-2:2023cdq}, the $W$ boson mass~\cite{CDF:2022hxs}, and the rare kaon~\cite{NA62:2024pjp, KOTO:2024zbl} and $B$ meson~\cite{Belle-II:2023esi} decays. Among numerous well-motivated new physics models, extensions to the gauge sector of the SM with an extra $U(1)$ gauge field have received considerable attention. It is usually introduced either through kinetic mixing in the case of the dark photon model~\cite{Fayet:1980ad, Fayet:1980rr, Holdom:1986eq, Okun:1982xi}, or anomaly free $U(1)’$ charges in the $Z’$ model~\cite{Fayet:1990wx, Leike:1998wr, Langacker:2008yv}. These hypothetical particles are also promising portals that could potentially serve as a bridge between the SM and the dark matter sector.

Although numerous experimental searches for the dark photon have been undertaken at $e^+ e^-$~\cite{BaBar:2014zli, BaBar:2017tiz} and hadron-hadron colliders~\cite{LHCb:2019vmc, CMS:2019buh}, as well as fixed target experiments~\cite{Banerjee:2019pds, Andreev:2021fzd}, no direct signal of a narrow resonance has been found so far. The strongest constraints on the dark photon parameters come from the BaBar~\cite{BaBar:2014zli, BaBar:2017tiz} and the CMS~\cite{CMS:2019buh} collaborations, leading to an upper limit of $\epsilon \le 10^{-3}$ over a wide range of the dark photon mass, albeit with a few gaps associated with resonance production. However, theoretical investigations on decay-agnostic processes in connection with electroweak precision observables~\cite{Hook:2010tw, Curtin:2014cca, Loizos:2023xbj} and electron-nucleon deep-inelastic scattering (DIS)~\cite{Kribs:2020vyk, Thomas:2021lub, Yan:2022npz} have led to much weaker constraints, with upper bounds on $\epsilon$ being of ${\cal O} (10^{-2}) \sim {\cal O}(10^{-1})$. While a recent global QCD analysis provided indirect evidence for the existence of a dark photon~\cite{Hunt-Smith:2023sdz}, the preferred range of values of the mixing parameter, $\epsilon \in (0.02, 0.14)$, 
is in tension with the constraints from direct searches.

It has been noticed that the constraints from direct experimental searches are model-dependent, and that these  could be significantly relaxed in light of the potential couplings of the dark boson to dark matter particles~\cite{Abdullahi:2023tyk, Queiroz:2024ipo}. As an example, in this work we examine the limits on the dark photon parameters which were set by the CMS collaboration~\cite{CMS:2019buh} under the assumption that the dark photon is a narrow resonance, because it only decays to SM particles. Those limits are re-assessed taking into account the impact of a larger decay width of the dark photon, resulting from its decay to light dark matter.

We begin with a brief review of the dark photon formalism in Sec.~\ref{sec:Formalism}. We present the cross section of $p p \to \mu^+ \mu^- X$ with the inclusion of a dark photon in Sec.~\ref{sec:cross-section}. The exclusion constraints on the mixing parameter for a broad dark photon are derived in Sec.~\ref{sec:Results}. Finally, we summarize our conclusions in Sec.~\ref{sec:Conclusion}.


\section{The Dark Photon Formalism}
\label{sec:Formalism}

The dark photon is typically introduced into the SM as an additional massive U(1) gauge boson which kinetically with the SM hypercharge $B$ boson~\cite{Fayet:1980ad, Fayet:1980rr, Holdom:1986eq, Okun:1982xi},
\begin{align}
    \label{eq:L_EW}
        \mathcal{L} \supset - &\frac{1}{4} F_{\mu \nu} F^{\mu \nu} - \frac{1}{4} \bar{Z}_{\mu \nu} \bar{Z}^{\mu \nu} + \frac{1}{2} m^2_{\bar{Z}} \bar{Z}_{\mu} \bar{Z}^{\mu} \nonumber\\
        & - \frac{1}{4} F'_{\mu \nu} F'^{\mu \nu} + \frac{1}{2} m^2_{A'} A'_{\mu} A'^{\mu}  + \frac{\epsilon}{2 \cos{\theta_W}} F'_{\mu \nu} B^{\mu \nu}\, .
    \end{align}
Here, $\bar{Z}$ and $A'$ denote the unmixed versions of the SM neutral weak boson and the dark photon, respectively. $F'_{\mu \nu}$ is the field strength tensor of the dark photon, and $\epsilon$ is the kinetic mixing parameter. The dark photon mass $m_{A'}$ could be generated from the dark Higgs~\cite{Galison:1983pa} or the Stueckelberg mechanism~\cite{Stueckelberg:1938hvi, Kors:2004dx}. The former case has richer phenomenological implications, though it is not so important in the present work as the (dark)Higgs contributions to $p p \to \mu^+ \mu^- X$ are negligibly small.

The physical $Z$ and dark photon $A_D$ can be derived by performing field redefinitions, i.e.
\begin{align}
Z_{\mu} & = \cos{\alpha} \bar{Z}_{\mu} + \sin{\alpha} A'_{\mu}\, ,\nonumber\\
{A_{D}}_{\mu} & = -\sin{\alpha} \bar{Z}_{\mu} + \cos{\alpha} A'_{\mu}\, ,
\end{align}
where $\alpha$ is the $\bar{Z} - A'$ mixing angle defined as~\cite{Kribs:2020vyk}
\begin{align}
\tan{\alpha} & = \frac{1}{2\epsilon_{W}} \bigg(1 - \epsilon_{W}^2 - \rho^2 \nonumber\\ 
& - \text{sign}(1 - \rho^2) \sqrt{4 \epsilon_{W}^2 + (1 - \epsilon_{W}^2 - \rho^2)^2} \bigg)\ ,
\end{align}
with
\begin{equation}
\epsilon_W = \frac{\epsilon \tan{\theta_W}}{\sqrt{1 - \epsilon^2 / \cos^2\theta_W}}\, ,\ \ 
\rho = \frac{m_{A'}/m_{\bar{Z}}} {\sqrt{1 - \epsilon^2 / \cos^2\theta_W}}\, .
\end{equation}

Diagonalising the resulting mass-squared matrix gives the masses of these physical states~\cite{Kribs:2020vyk},
\begin{eqnarray}
\label{eq:m_Z_AD}
M^2_{Z, A_D} &=& \frac{m_{\bar{Z}}^2}{2} [ 1 + \epsilon_W^2 + \rho^2 \nonumber\\
&& \pm {\rm sign}(1-\rho^2) \sqrt{(1 + \epsilon_W^2 + \rho^2)^2 - 4 \rho^2} ] \, .
\end{eqnarray}

The weak couplings of the physical $Z$ boson (in unit of $e=\sqrt{4\pi\alpha_{\rm em}}$) will deviate from their SM values~\cite{Kribs:2020vyk},
\begin{align}
C_Z^v &= (\cos\alpha - \epsilon_W \sin\alpha) C_{\bar Z}^v + \epsilon_W \sin\alpha \cot \theta_W C_{\gamma}^v ,\nonumber\\
C_Z^a &= (\cos\alpha - \epsilon_W \sin\alpha) C_{\bar Z}^a \, ,
\end{align}
whilst the dark photon couplings to SM fermions are 
\begin{align}
C^{v}_{A_D} &= -(\sin{\alpha} + \epsilon_W \cos{\alpha}) C^v_{\bar{Z}} + \epsilon_{W} \cos{\alpha}\cot\theta_W C^v_{\gamma}\, ,\nonumber\\
C^{a}_{A_D} & = -(\sin{\alpha} + \epsilon_W \cos{\alpha}) C^a_{\bar{Z}}\, ,
\end{align}
where $C^v_{\bar{Z}}$ and $C^a_{\bar{Z}}$ are the SM weak couplings,
\begin{align}
& \big\{
C^v_{\bar{Z},e}, C^v_{\bar{Z},u}, C^v_{\bar{Z},d} 
\big\}\, \sin 2\theta_W \nonumber\\
&= 
\Big\{ 
- \frac{1}{2} + 2 \sin^2\theta_W,\,
  \frac{1}{2} - \frac{4}{3}\sin^2\theta_W,\,
- \frac{1}{2} + \frac{2}{3}\sin^2\theta_W 
\Big\}\, ,\nonumber
\\
& \big\{
C^a_{\bar{Z},e}, C^a_{\bar{Z},u}, C^a_{\bar{Z},d} 
\big\}\, \sin 2\theta_W
= \Big\{ - \frac12,\, \frac12,\, - \frac12 \Big\}\, , 
\end{align}
and $C^v_{\gamma} = \{ C^v_{\gamma,e}, C^v_{\gamma,u}, C^v_{\gamma, d}\} = \{-1, 2/3, -1/3\}$ are the electromagnetic couplings. For abbrevity, we will define
\bea
C^v_{Z,f} &=& v_f\, ,\ \ C^a_{Z,f} = a_f\, ,\nonumber\\
C^v_{A_D,f} &=&  \bar{v}_f\, ,\ \ C^a_{A_D,f} = \bar{a}_f\, .
\eea

As one of the promising portals connecting to the dark sector, the dark photon could also couple to dark matter particles. Popular scenarios include Dirac, pseudo-Dirac, scalar and asymmetric dark matter models~\cite{Izaguirre:2015yja}. 
In this work, we aim to investigate the impact a larger decay width of the dark photon has on the limits of the kinetic mixing parameter, regardless of the specific type of dark matter particle.
That being said, for simplicity, we consider Dirac fermion dark matter, $\chi$, with the interaction~\cite{Loizos:2023xbj} 
\begin{equation}
{\cal L}_{\chi} = g_{\chi} \bar{\chi} \gamma^{\mu} \chi A'_{\mu}\, ,
\end{equation}
and the physical coupling
\be
C_{A_D, \chi\bar{\chi}} = \frac{g_{\chi} \cos\alpha}{\sqrt{1- \epsilon^2/\cos^2\theta_W}}\, .
\ee
Using these couplings, the total decay width of the dark photon is~\cite{Wang:2025rsg}
\begin{align}
\label{eq:AD_Total_Width}
    \Gamma_{A_D} = \Gamma_{A_D \to {\rm SM}} + \Gamma_{A_D \to \chi\bar{\chi}}
\end{align}
where
\begin{align}
\label{eq:Gamma-AD}
\Gamma_{A_D \to {\rm SM}} = & \sum_{f} N_C^f \cdot \frac{M_{A_D} \alpha_{\rm em}}{3} \bigg\{\left(1 + \frac{2m_{f}^2}{M_{A_D}^2}\right) \bar{v}_{f}^2 + \nonumber\\
& \left(1 - \frac{4m_{f}^2}{M_{A_D}^2}\right)\bar{a}_{f}^2\bigg\}\sqrt{1 - \frac{4m_{f}^2}{M_{A_D}^2}}\, ,\nonumber\\
\Gamma_{A_D \to \chi\bar{\chi}} = & \frac{M_{A_D} C^2_{A_D,\chi\bar{\chi}}}{12\pi} \left(1 + \frac{2m_{\chi}^2}{M_{A_D}^2}\right) \sqrt{1 - \frac{4m_{\chi}^2}{M_{A_D}^2}}\, ,
\end{align}
with $N_C^f = 1$ for leptons and $N_C^f = 3$ for quarks. The coupling $g_{\chi}$ is typically of ${\cal O}(1)$, therefore, $\Gamma_{A_D \to \chi\bar{\chi}}$ will be a few orders of magnitude larger than $\Gamma_{A_D \to \rm SM}$ which is suppressed by $\epsilon^2$.\\


\section{Cross section of $p p \to \mu^+\mu^- X$}
\label{sec:cross-section}

\begin{figure}[!h]
    \begin{center}
        \includegraphics[width=0.9\columnwidth]{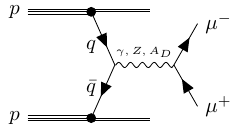}
        \vspace*{-0.2cm}
        \caption{The $pp \to \mu^+ \mu^- X$ process with the inclusion of a dark photon.}
        \label{fig:feynman}
    \end{center}
    \end{figure}

The $pp \to \mu^+ \mu^- X$ process with the inclusion of a dark photon is depicted in Fig.~\ref{fig:feynman}. In the center-of-mass frame, the momenta of the initial protons can be defined as
\begin{align}
P_1 &= \frac{\sqrt{s}}{2} (1,0,0,1)\, ,\nonumber\\
P_2 &= \frac{\sqrt{s}}{2} (1,0,0,-1)\, ,
\end{align}
where we have neglected the proton mass.

\subsection{Partonic cross section}
In the center-of-mass frame of $q (p_1) \bar{q}(p_2) \to \mu^- (k_1) \mu^+ (k_2)$ scattering, we use a Breit-Wigner parametrization for the $s$-channel $Z$ boson and dark photon scattering amplitudes~\cite{Cahn:1986qf} with invariant mass
\begin{equation}
\hat{s} = (p_1 + p_2)^2\, .
\end{equation}
The total partonic cross section can be written as 
\begin{equation}
\sigma_{q\bar{q}\to \mu^-\mu^+} = \sigma_{\gamma} + \sigma_{Z} + \sigma_{\gamma Z}  + \sigma_{A_D} + \sigma_{\gamma A_D} + \sigma_{Z A_D}\, .
\end{equation}
The individual terms read
\begin{align}
\sigma_{\gamma} &= \frac{4\pi \alpha^2_{\rm em}}{3 \hat{s}} \cdot Q^2_q\, ,\nonumber\\
\sigma_{Z} &= \frac{4\pi \alpha^2_{\rm em}}{3 \hat{s}}  \cdot (v^2_q + a^2_q) (v^2_{\mu} + a^2_{\mu}) \cdot \frac{\hat{s}^2}{(\hat{s} - M^2_Z)^2 + \hat{s} \Gamma^2_Z(\hat{s})}\, ,\nonumber\\
\sigma_{\gamma Z} &= \frac{4\pi \alpha^2_{\rm em}}{3 \hat{s}} \cdot v_q v_{\mu} \cdot \frac{- 2 Q_q \cdot \hat{s} (\hat{s} - M^2_Z)}{(\hat{s} - M^2_Z)^2 + \hat{s} \Gamma^2_Z(\hat{s})} \, ,\nonumber\\
\sigma_{A_D} &= \frac{4\pi \alpha^2_{\rm em}}{3 \hat{s}} (\bar{v}^2_q + \bar{a}^2_q) (\bar{v}^2_{\mu} + \bar{a}^2_{\mu}) \frac{\hat{s}^2}{( \hat{s} - M^2_{A_D})^2 + \hat{s} \Gamma^2_{A_D}(\hat{s})} \, ,\nonumber\\
\sigma_{\gamma A_D} &= \frac{4\pi \alpha^2_{\rm em}}{3 \hat{s}} \cdot \bar{v}_q \bar{v}_{\mu} \cdot \frac{- 2 Q_q \cdot \hat{s} (\hat{s} - M^2_{A_D})}{(\hat{s} - M^2_{A_D})^2 + \hat{s} \Gamma^2_{A_D}(\hat{s})} \, ,\nonumber\\
\sigma_{Z A_D} &= \frac{4\pi \alpha^2_{\rm em}}{3 \hat{s}} \cdot (v_q \bar{v}_q + a_q \bar{a}_q) (v_{\mu} \bar{v}_{\mu} + a_{\mu} \bar{a}_{\mu}) \nonumber\\
& \cdot \frac{2 \hat{s}^2 [(\hat{s} - M^2_Z) (\hat{s} - M^2_{A_D}) + \hat{s} \Gamma_Z(\hat{s}) \Gamma_{A_D}(\hat{s})]}{ [(\hat{s} - M^2_Z)^2 + \hat{s} \Gamma^2_Z(\hat{s})] [(\hat{s} - M^2_{A_D})^2 + \hat{s} \Gamma^2_{A_D}(\hat{s})]} \, ,
\end{align}
where $Q_q$ is the electric charge of the quark $q$, and $\Gamma_{Z(A_D)}(\hat{s}) = \sqrt{\hat{s}} \Gamma_{Z(A_D)} / M_{Z(A_D)}$.\\

\subsection{Hadronic cross section}
The partonic scattering cross section is converted into the cross section for the hadronic process by taking the convolution of the partonic scattering cross section with the appropriate parton distribution functions (PDFs)~\cite{Kenyon:1982tg}
\begin{align}
\label{eq:diff_cross_section}
    \frac{d\sigma_{pp \to \mu^{-}\mu^{+}X}}{d m^2_{\mu\mu}} &= \frac{1}{3} \sum_{q} \frac{1}{s} \sigma_{q\bar{q} \to \mu^-\mu^+}(m^2_{\mu\mu}) \cdot \nonumber\\
    &\int_{\tau}^{1} \frac{d x_1}{x_1}
    \Big[f_{q}(x_1,Q^2)f_{\bar{q}}(\tau/x_1,Q^2) + \{ q \leftrightarrow \bar{q} \} \Big]\, ,
\end{align}
where $m_{\mu\mu}$ is the di-muon invariant mass. $f_q$ ($f_{\bar q}$) is the quark (anti-quark) distribution function in the proton, with $\tau = \hat{s}/s$. The pre-factor $1/3$ is due to the fact that the quark and anti-quark color must match.

When comparing with the experimental data, one should also take into account the finite energy resolution of the detector, which can be done by convoluting the differential cross section in Eq.~(\ref{eq:diff_cross_section}) with the following Gaussian smearing function~\cite{BeamBroadening1,Jadach:2015cwa}, 
\begin{align}
\label{eq:Gaussian}
    G(m'_{\mu\mu}, m_{\mu\mu}) = \frac{1}{\delta_{res} \sqrt{2\pi}} e^{-\frac{1}{2}\frac{(m'_{\mu\mu} - m_{\mu\mu})^2}{\delta_{res}^2}}\, ,
\end{align}
where $\delta_{res}$ is taken to be $2\%$ of the partonic center-of-mass energy $\sqrt{\hat{s}} = m_{\mu\mu}$~\cite{CMS:2018mdl}.

\section{Constraints on the dark photon parameters}
\label{sec:Results} 

In our numerical analysis, we use $\alpha^{-1}_{\rm em}(M^2_Z) = 128.952$ and $\sin^2\theta_W (M^2_Z)|_{\overline{\rm MS}} = 0.23129$~\cite{ParticleDataGroup:2024cfk}. For the parton distributions, we take the next-to-leading order (NLO) results with $\alpha_s(M^2_Z) = 0.118$ from the NNPDF collaboration~\cite{NNPDF:2021njg}. 

Currently, constraints on $g_{\chi}$ associated with the dark matter relic density and direct detection are usually expressed in terms of the dimensionless combination~\cite{Izaguirre:2015yja, Filippi:2020kii, Krnjaic:2025noj}
\begin{equation}
y = \epsilon^2 \alpha_D \left( \frac{m_{\chi}}{m_{A'}} \right)^4\, ,
\end{equation}
with $\alpha_D = g_{\chi}^2 / 4\pi$ varying from $\alpha_{\rm em}$ to the perturbativity bound, $\alpha_D = 1/2$. 
However, these analyses have focused on the light mass region within an alternative model in which the dark photon kinetically mixes with the SM photon, i.e. ${\cal L}_{\rm mix} = \epsilon F'_{\mu\nu} F^{\mu\nu}$. Dedicated constraints on the dark parameters in the framework of the model used in this analysis, that is based upon Eq.~(\ref{eq:L_EW}), are still lacking, especially in the heavy mass region.

A recent analysis of the electroweak precision observables based on Eq.~(\ref{eq:L_EW}) has attempted to set upper bounds directly on $g_{\chi}$~\cite{Loizos:2023xbj}. However, the electroweak observables lose sensitivity to $g_{\chi}$ when $\epsilon$ approaches its upper limits. As a result, $g_{\chi}$ can vary over a wide range, up to the perturbativity bound. In the following, we will take typical values of $g_{\chi}$ in the conservative range of $\alpha_{D} < 0.1$.
\begin{figure}[h]
    \begin{center}
        \includegraphics[width=\columnwidth]{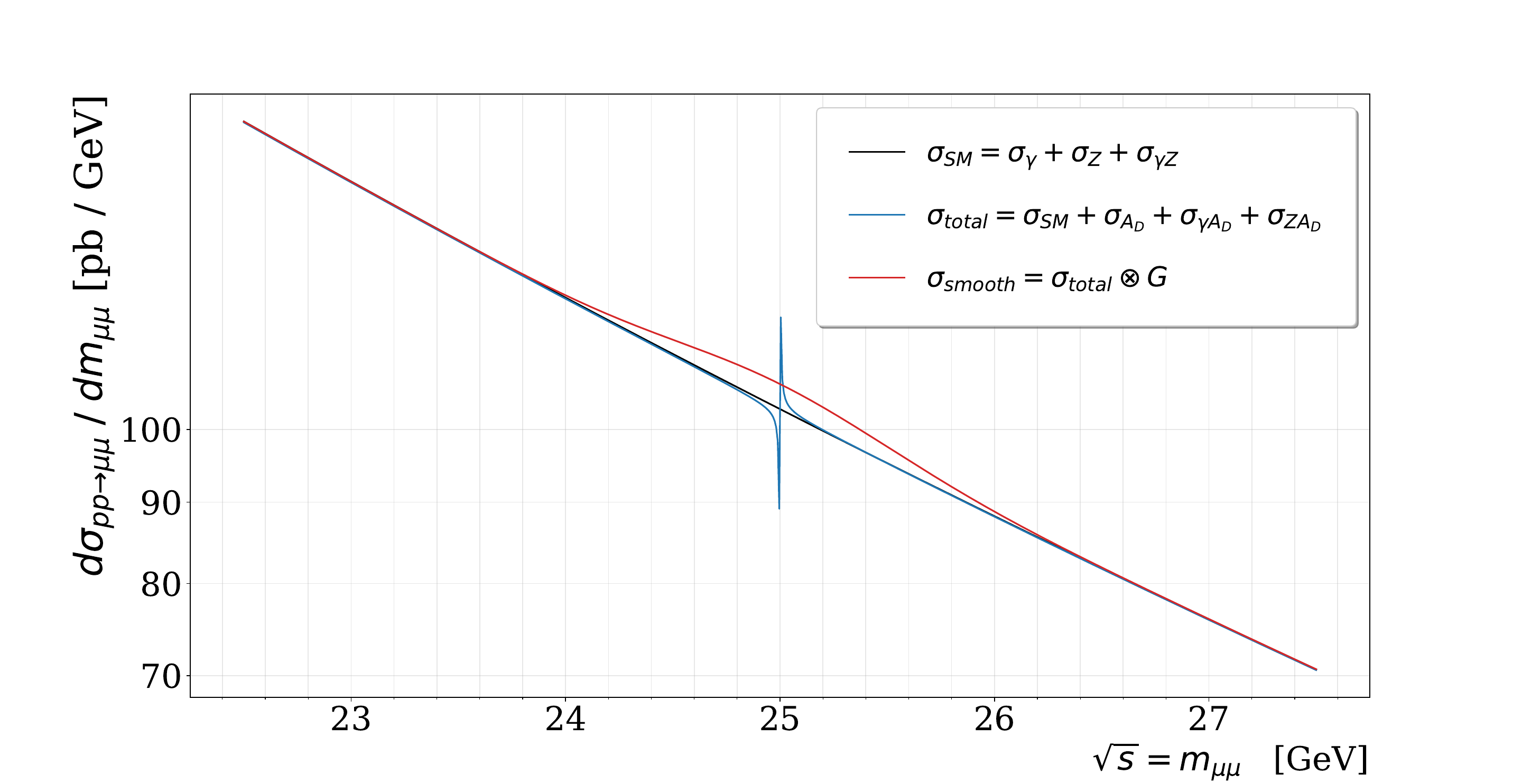}
        \vspace*{-0.2cm}
        \caption{The signal for a narrow $(\Gamma_{A_D} = 8.9\ {\rm keV})$ dark photon with $(M_{A_D}, \epsilon^2, g_{\chi}) = (25\ {\rm GeV}, 2\times10^{-5}, 0)$. The signal before and after the inclusion of the dark photon is in black and blue respectively. The resulting convolution of the blue signal with the Gaussian smearing function in Eq.~(\ref{eq:Gaussian}) is in red.}
        \label{fig:AD_signal_gx_0}
\end{center}
\end{figure}

\begin{figure}[h]
\begin{center}
    \includegraphics[width=\columnwidth]{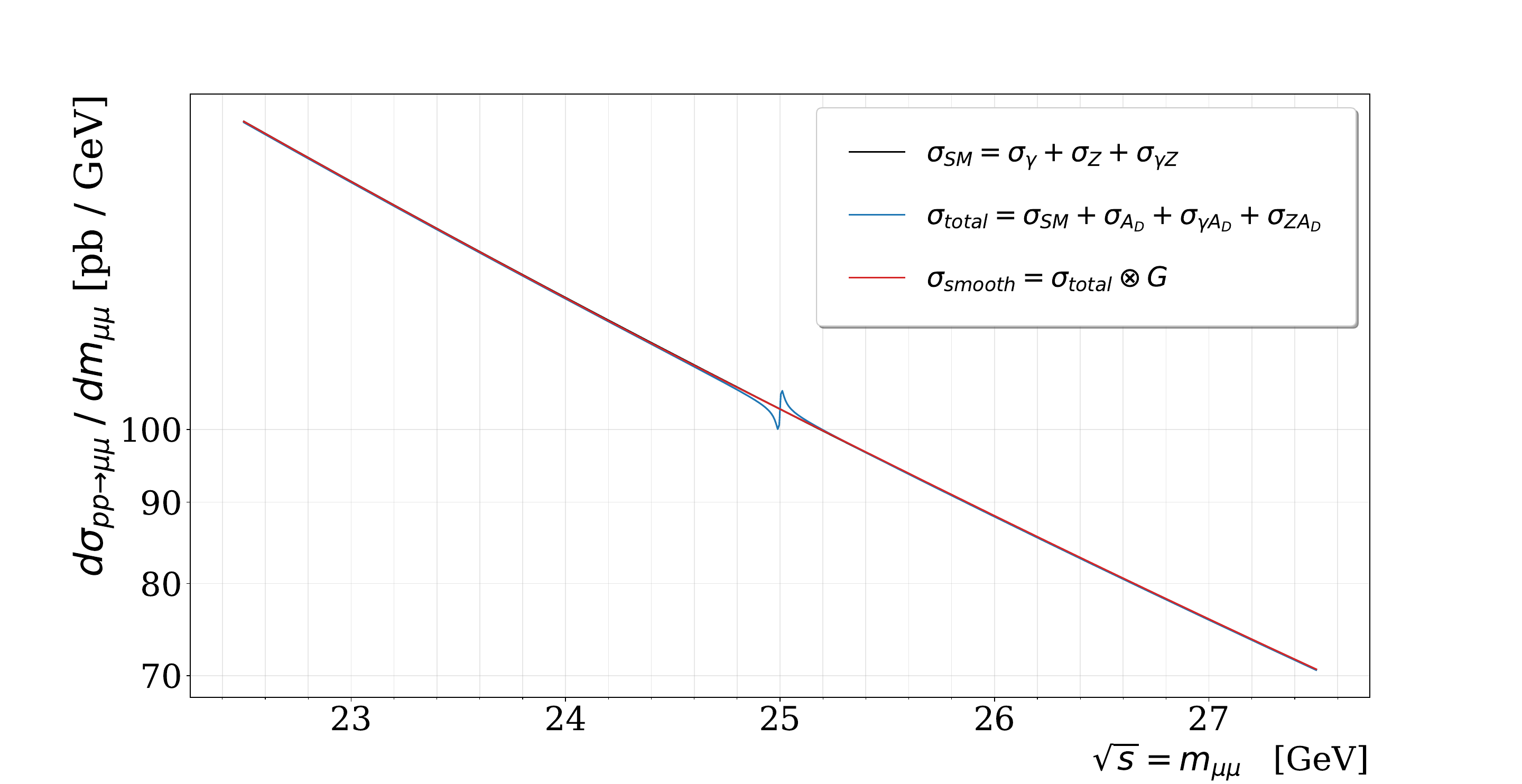}
    \vspace*{-0.2cm}
    \caption{The signal for a broad $(\Gamma_{A_D} = 0.016\ {\rm GeV})$ dark photon with $(M_{A_D}, \epsilon^2, g_{\chi}) = (25\ {\rm GeV}, 2\times10^{-5}, 0.05)$. The definition of the black, blue and red signals are the same as in Fig.~\ref{fig:AD_signal_gx_0}.}
    \label{fig:AD_signal_gx_0p05a}
\end{center}
\end{figure}
\begin{figure}[h]
\begin{center}
    \includegraphics[width=\columnwidth]{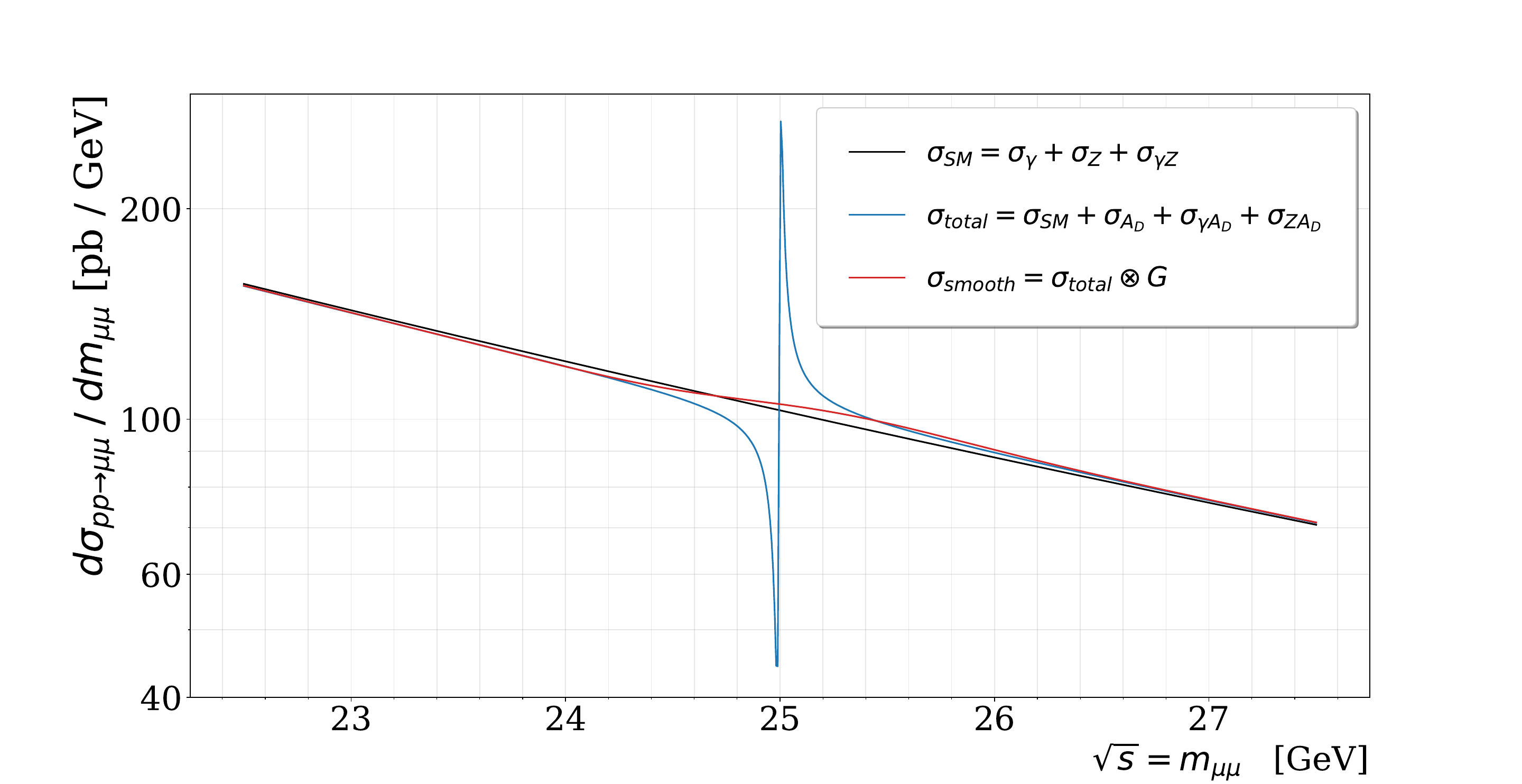}
    \vspace*{-0.2cm}
    \caption{The signal for a broad $(\Gamma_{A_D} = 0.016\ {\rm GeV})$ dark photon with $(M_{A_D}, \epsilon^2, g_{\chi}) = (25\ {\rm GeV}, 6.6\times10^{-4}, 0.05)$. The definition of the black, blue and red signals are the same as in Fig.~\ref{fig:AD_signal_gx_0}.}
    \label{fig:AD_signal_gx_0p05}
\end{center}
\end{figure}

The upper limits on the mixing parameter set by the CMS collaboration~\cite{CMS:2019buh}, $\epsilon^{\rm CMS}$, were derived by assuming that the dark photon only decays to SM particles, i.e., $g_{\chi} = 0$. In this case, a narrow resonance signal is expected in the di-muon invariant mass spectrum. For example, by taking $M_{A_D} = 25\ {\rm GeV}$ and $\epsilon^2 = 2\times 10^{-5}$~\cite{CMS:2019buh}, we have
\begin{equation}
\Gamma_{A_D} = \Gamma_{A_D \to {\rm SM}} = 8.9\ {\rm keV}\, .
\end{equation}
With these parameters, the dark photon cross sections are shown in Figs.~\ref{fig:AD_signal_gx_0}. The asymmetrical behaviour of the blue signal originates from the $\gamma - A_D$ interference term, although once the energy resolution of the detector is accounted for the $\gamma - A_D$ term is suppressed and the signal is smoothed out, as seen in red.

When $g_{\chi}$ is turned on, the total decay width is dominated by the $\chi\bar{\chi}$ channel. For example, take $g_{\chi} = 0.05$, then 
\begin{equation}
\Gamma_{A_D} \sim \Gamma_{A_D \to \chi\bar{\chi}} (g_{\chi} = 0.05) = 0.016\ {\rm GeV}\, ,
\end{equation}
which will result in a considerably broader resonance signal, thus significantly reducing its visibility, as shown in Figs.~\ref{fig:AD_signal_gx_0p05a} and \ref{fig:AD_signal_gx_0p05}. To recover the smoothed-out (red) signal size seen in Fig.~\ref{fig:AD_signal_gx_0}, a larger value of $\epsilon$ is required. In this case, it was found, using Eq.~(\ref{eq:match}), that increasing $\epsilon^2$ to  $6.6\times10^{-4}$ successfully reproduces the signal size. However, it still maintains the asymmetrical profile seen in the blue signal. Once again, this asymmetry originates from the $\gamma - A_D$ interference term, although, unlike in Fig.~\ref{fig:AD_signal_gx_0}, accounting for the energy resolution of the detector no longer suppresses this term, because of the increase in $\Gamma_{A_D}$.

We can now re-evaluate the upper limit on $\epsilon$, with non-zero values of $g_{\chi}$, by equating the maximum peak height (relative to the SM background) between the narrow $(g_{\chi} = 0)$ and broad $(g_{\chi} > 0)$ resonances. As an example, taking the upper limits on $\epsilon$ set by the CMS collaboration~\cite{CMS:2019buh}, we require the following matching condition,
\begin{align}
\label{eq:match}
    &{\rm max}\left| \frac{d\sigma_{\rm smooth}}{d m^2_{\mu\mu}} (\epsilon, M_{A_D}, g_{\chi}) - \frac{d\sigma_{\rm SM}}{d m^2_{\mu\mu}}\right| = \notag\\ 
    &{\rm max}\left| \frac{d\sigma_{\rm smooth}}{d m^2_{\mu\mu}} (\epsilon^{\rm CMS}, M_{A_D}, g_{\chi} = 0) - \frac{d\sigma_{\rm SM}}{d m^2_{\mu\mu}}\right| \, .
\end{align}

\begin{figure*}
\begin{center}
    \includegraphics[width=\textwidth]{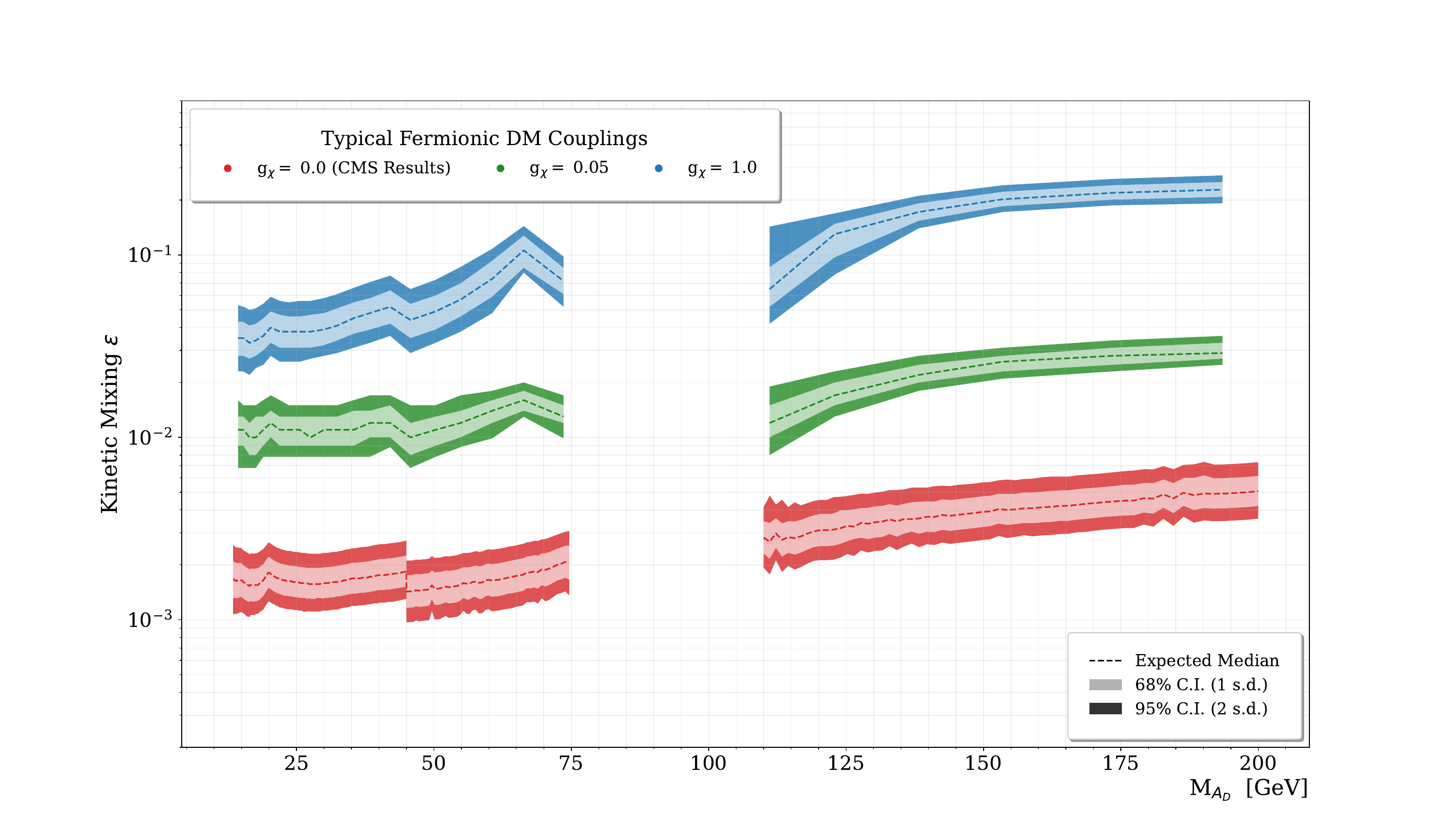}
    \vspace*{-0.2cm}
    \caption{ Here we have the $90\%$ C.L. median expected limit on $\epsilon$ (dashed) with the corresponding $68\%$ (light shaded) and $95\%$ (dark shaded) confidence intervals. The limits set in \cite{CMS:2019buh} are in red and the qualitative relaxed limits for $g_{\chi} = 0.05$ and $g_{\chi} = 1$ are in green and blue respectively.}
    \label{fig:relaxed_limits}
\end{center}
\end{figure*}

The results are shown in Fig.~\ref{fig:relaxed_limits}. With reasonable choices of the coupling $g_{\chi}$,  satisfying the constraints set in Refs.~\cite{Filippi:2020kii, Loizos:2023xbj}, the upper bounds on $\epsilon$ will be relaxed by one to two orders of magnitude compared with the CMS constraints. 

Even with a small coupling, $g_{\chi} =
0.05$, the resulting exclusion limits will be shifted to ${\cal O}(10^{-2})$, which are comparable with the results determined from electroweak precision observables~\cite{Curtin:2014cca, Loizos:2023xbj}. For $g_{\chi} = 1.0$, the upper limits of $\epsilon$ will be as large as ${\cal O}(10^{-1})$, consistent with those from electron-proton deep-inelastic scattering (DIS)~\cite{Kribs:2020vyk, Thomas:2021lub, Yan:2022npz}.\\

Taking the dark parameters at the lower end of of the top band in Fig.~\ref{fig:relaxed_limits}, with $g_{\chi}=1.0$, $\epsilon=0.04$ and fixing $m_{\chi}/m_{A'} = 1/3$, we find $y=1.6\times 10^{-6}$, which is well above the lower bounds set by relic density in various models~\cite{Izaguirre:2015yja}. As a result, the dark matter, $\chi$, only constitutes part of the total relic abundance. In this case the current upper limits on $y$ from direct detection, which were derived by assuming that $\chi$ accounts for all of the dark matter, should be relaxed. 

%


\section{Conclusion}
\label{sec:Conclusion}

We have calculated the cross section of the $p p \to \mu^- \mu^+ X$ process by including the contributions from a dark photon. While a narrow resonance is expected if the dark photon only decays to the Standard Model particles, its total decay width could be much larger if it were to also couple to dark matter particles. We investigated the impact of a larger decay width on the exclusion limits of the dark photon parameters, by revisiting the constraints set by the CMS collaboration. 

With typical values of the coupling $g_{\chi}$, the upper bounds on the mixing parameter,  $\epsilon$, will be significantly relaxed to be consistent with the determinations from electroweak precision observables and even  electron-nucleon deep-inelastic scattering.

The analysis presented in this work can also be applied to direct experimental searches for other forms of new gauge bosons. Moreover, we would suggest that future direct experimental searches at $e^+ e^-$ and hadron colliders, as the most promising avenues for discovery, could put more emphasis on the need to analyze signals associated with the production of a broad dark boson resonance.


\section*{Acknowledgments}
This work was supported by the University of Adelaide and by the Australian Research Council through the Centre of Excellence for Dark Matter Particle Physics (CE200100008). The funding for J.~R.~Felix was provided by an Australian Government Research Training Program Scholarship and the Cedric Arnold Seth Ferry Scholarship.


\bibliography{bibliography}
\end{document}